\documentclass[letter]{aa}
\usepackage[varg]{txfonts}
\usepackage{graphicx}
\usepackage{natbib,twoopt}
\usepackage[breaklinks=true]{hyperref} 
\usepackage{lscape}
\bibpunct{(}{)}{;}{a}{}{,} 
\makeatletter
 \newcommandtwoopt{\citeads}[3][][]{\href{https://ui.adsabs.harvard.edu/abs/#3/abstract}%
 {\def\hyper@linkstart##1##2{}%
 \let\hyper@linkend\@empty\citealp[#1][#2]{#3}}}
 \newcommandtwoopt{\citepads}[3][][]{\href{https://ui.adsabs.harvard.edu/abs/#3/abstract}%
 {\def\hyper@linkstart##1##2{}%
 \let\hyper@linkend\@empty\citep[#1][#2]{#3}}}
 \newcommandtwoopt{\citetads}[3][][]{\href{https://ui.adsabs.harvard.edu/abs/#3/abstract}%
 {\def\hyper@linkstart##1##2{}%
 \let\hyper@linkend\@empty\citet[#1][#2]{#3}}}
 \newcommandtwoopt{\citeyearads}[3][][]%
 {\href{https://ui.adsabs.harvard.edu/abs/#3/abstract}
 {\def\hyper@linkstart##1##2{}%
 \let\hyper@linkend\@empty\citeyear[#1][#2]{#3}}}
\makeatother

\begin{document}

   \title{Mini-moons from horseshoes: A physical characterization of 2022~NX$_{1}$ with OSIRIS 
          at the 10.4~m Gran Telescopio Canarias\thanks{Based on observations made with the 
          Gran Telescopio Canarias (GTC) telescope, in the Spanish Observatorio del Roque de 
          los Muchachos of the Instituto de Astrof\'{\i}sica de Canarias (program ID GTC23-22A).}}
   \author{R.~de~la~Fuente Marcos\inst{1}
            \and
           J.~de~Le\'on\inst{2,3}
            \and 
           C.~de~la~Fuente Marcos\inst{4}
            \and
           J. Licandro\inst{2,3}
            \and
           M. Serra-Ricart\inst{2,3}
            \and
           A. Cabrera-Lavers\inst{5,2,3}
          }
   \authorrunning{R. de la Fuente Marcos et al.}
   \titlerunning{Physical characterization of 2022~NX$_{1}$ with GTC} 
   \offprints{R. de la Fuente Marcos, \email{rauldelafuentemarcos@ucm.es}}
   \institute{AEGORA Research Group,
              Facultad de Ciencias Matem\'aticas,
              Universidad Complutense de Madrid,
              Ciudad Universitaria, E-28040 Madrid, Spain
              \and
              Instituto de Astrof\'{\i}sica de Canarias (IAC),
              C/ V\'{\i}a L\'actea s/n, E-38205 La Laguna, Tenerife, Spain
              \and
              Departamento de Astrof\'{\i}sica, Universidad de La Laguna,
              E-38206 La Laguna, Tenerife, Spain
              \and
              Universidad Complutense de Madrid,
              Ciudad Universitaria, E-28040 Madrid, Spain
              \and
              GRANTECAN,
              Cuesta de San Jos\'e s/n, E-38712 Bre\~na Baja, La Palma, Spain
             }
   \date{Received 21 November 2022 / Accepted 21 January 2023}
   \abstract
      {The near-Earth orbital space is shared by natural objects and space 
       debris that can be temporarily captured in geocentric orbits. 
       Short-term natural satellites are often called mini-moons. Reflectance 
       spectroscopy can determine the true nature of transient satellites 
       because the spectral signatures of spacecraft materials and near-Earth 
       asteroids (NEAs) are different. The recently discovered object 
       2022~NX$_{1}$ follows an Earth-like orbit that turns it into a 
       recurrent but ephemeral Earth companion. It has been suggested that 
       2022~NX$_{1}$ could have an artificial origin or be lunar ejecta. 
       }
      {Here, we use reflectance spectroscopy and $N$-body simulations to 
       determine the nature and actual origin of 2022~NX$_{1}$.
       }
      {We carried out an observational study of 2022~NX$_{1}$, using the 
       OSIRIS camera spectrograph at the 10.4~m Gran Telescopio Canarias, to 
       derive its spectral class. $N$-body simulations were also performed to 
       investigate how it reached NEA space.
       }
      {The reflectance spectrum of 2022~NX$_{1}$ is neither compatible with 
       an artificial origin nor lunar ejecta; it is also different from the 
       V type of the only other mini-moon with available spectroscopy, 
       2020~CD$_{3}$. The visible spectrum of 2022~NX$_{1}$ is consistent 
       with that of a K-type asteroid, although it could also be classified 
       as an Xk type. Considering typical values of the similar albedo of 
       both K-type and Xk-type asteroids and its absolute magnitude, 
       2022~NX$_{1}$ may have a size range of 5 to 15~m. We confirm that 
       2022~NX$_{1}$ inhabits the rim of Earth's co-orbital space, the 1:1 
       mean-motion resonance, and experiences recurrent co-orbital 
       engagements of the horseshoe-type and mini-moon events. 
       }
      {The discovery of 2022~NX$_{1}$ confirms that mini-moons can be larger 
       than a few meters and also that they belong to a heterogeneous 
       population in terms of surface composition.
       }


   \maketitle

   \section{Introduction\label{Intro}}
      The Moon is Earth's only permanent natural satellite but over 22,000 artificial objects (active spacecraft and space debris) of all 
      sizes also orbit our planet \citep{GCAT}.\footnote{\href{https://planet4589.org/space/gcat/}{https://planet4589.org/space/gcat/}} In 
      addition, passing bodies may be captured in geocentric orbits if they move at very low relative velocity inside the Hill radius of 
      Earth, 0.0098~AU; these include both small natural bodies \citep{2012Icar..218..262G} and hardware originally inserted in cislunar or 
      interplanetary space \citep{neosst1}.\footnote{\href{https://conference.sdo.esoc.esa.int/proceedings/neosst1/paper/470}
      {https://conference.sdo.esoc.esa.int/proceedings/neosst1/paper/470}} Natural temporarily captured orbiters of Earth or mini-moons 
      appear to be rare, difficult to spot objects \citep{{2020AJ....160..277F}}. Prior to 2022, only three small natural bodies had been 
      identified crossing into the region defined by negative geocentric energy. Most captured objects are eventually confirmed as returning 
      space debris.

      Reflectance spectroscopy can help to determine the true nature of transient satellites because the spectral signatures of spacecraft 
      materials and rocky asteroids are different. This technique is routinely used to identify space debris unambiguously (see for example 
      \citealt{2000PhDT........43J,2004AdSpR..34.1021J,2007A&ARv..14...41S,2017AdSpR..59.2488V,2021JAnSc..68.1186C}). 

      Spectral observations led to confirm that J002E3, an object found orbiting Earth in 2002, was the upper S-IVB stage of Apollo~12 
      which launched on November 14, 1969 \citep{2003DPS....35.3602J}. Low-resolution spectroscopy was also used to confirm that WT1190F, an 
      object that may have orbited our planet from 1998 until it impacted Earth on November 13, 2015, was space debris 
      \citep{2018Icar..304....4M,2019AdSpR..63..371B}, likely the translunar injection module of Lunar Prospector \citep{Watson2016}. So 
      far and out of three small bodies identified as temporarily captured, only one, 2020~CD$_{3}$, has been studied spectroscopically 
      \citep{2020ApJ...900L..45B}.  

      The recently discovered object 2022~NX$_{1}$ \citep{2022MPEC....O...04B} follows an Earth-like orbit that turns it into a recurrent 
      but ephemeral Earth companion \citep{2022RNAAS...6..160D}. It has been suggested that 2022~NX$_{1}$ could have an artificial origin or 
      be lunar ejecta \citep{2022MPEC....O...04B}. Here, we use reflectance spectroscopy and $N$-body simulations to determine the nature 
      and actual origin of 2022~NX$_{1}$. This Letter is organized as follows. In Sect.~\ref{Data}, we introduce the context of our 
      research, review our methodology, and present the data and tools used in our analyses. In Sect.~\ref{Results}, we apply our 
      methodology to find out if 2022~NX$_{1}$ is natural or artificial and determine its probable origin. In Sect.~\ref{Discussion}, we 
      discuss our results. Our conclusions are summarized in Sect.~\ref{Conclusions}.

   \section{Context, methods, and data\label{Data}}
      In the following, we review some background material of theoretical nature needed to understand the results presented in the sections. 
      Basic details of our approach and the data are also included here as well as a summary of the tools used to obtain the results.

      \subsection{Dynamics background}
         Earth-approaching objects may remain in its vicinity following geocentric orbits as captured satellites, when the value of the 
         geocentric energy is negative \citep{1979RSAI...22..181C}. In addition, they could be subjected to resonant behavior and become 
         Earth co-orbitals trapped inside the 1:1 mean-motion resonance but following heliocentric paths, when the relative mean longitude 
         of the object with respect to Earth ($\lambda_{\rm r}$) oscillates about a fixed value \citep{2002Icar..160....1M}. However, most 
         visitors are just passing through and they are neither gravitationally captured by Earth nor engaged in the 1:1 orbital resonance 
         with our planet. 

         Here and in order to classify capture events, we follow the terminology discussed by \citet{2017Icar..285...83F}: an object that 
         does not complete at least one full revolution around Earth when bound is subjected to a temporarily captured flyby, but if it
         manages to complete at least one then we speak of a temporarily captured orbiter. On the other hand, if the value of 
         $\lambda_{\rm r}$ oscillates about 180{\degr}, with an amplitude $>$$\pi$, the object follows a horseshoe trajectory with respect 
         to Earth \citep{1999ssd..book.....M}.

      \subsection{Methodology}
         Reflectance spectroscopy requires the observation of a target object (natural or artificial) and one or more well-studied solar 
         analog stars at the same airmass as that of the object. The spectrum of the target is divided by the spectrum of the solar analog 
         (by each one if two or more and the resulting spectra are averaged) to produce the final reflectance spectrum of the object under 
         study. The entire data reduction process is described, for example, by \cite{2019A&A...625A.133L} and it consists of bias and 
         flat-field correction, background subtraction and extraction of the 1D spectrum from 2D images, and wavelength calibration. 

         The assessment of the past and future orbital evolution of an object and of its current dynamical state should be based on the 
         analysis of results from a representative sample of $N$-body simulations that take the uncertainties in the orbit determination 
         into account (see, for example, \citealt{2018MNRAS.473.2939D,2020MNRAS.494.1089D}). The near-Earth orbital domain is shaped by both 
         mean-motion and secular resonances that may lead to a chaotic dynamical evolution (see, for example, \citealt{2012Icar..217..355G}) 
         for both natural bodies and space debris even if the objects involved do not experience deep close encounters with the Earth-Moon 
         system and perhaps other planets. Statistical interpretation of the results is required if the evolution of the objects is 
         unstable. If the quality of the orbit determination is not sufficiently robust and if the dynamical evolution is chaotic, 
         predictions may only be reliable within a few decades (forward and backward in time) of the reference epoch.

      \subsection{Data, data sources, and tools}
         Object xkos033 was first observed by G.~Duszanowicz and J.~Camarasa using a 0.35-m, f/7.7 reflector telescope + CCD at Moonbase 
         South Observatory in the Hakos mountains, Namibia on July 2, 2022; fifteen days later, it was announced with the provisional 
         designation 2022~NX$_{1}$ \citep{2022MPEC....O...04B}. The discovery Minor Planet Electronic Circular (MPEC) states that ``The 
         Earth-like orbit of the object and its orbital evolution suggest that it could be of an artificial origin, launched from the Earth 
         decades ago or a lunar 
         ejecta.''\footnote{\href{https://minorplanetcenter.net/mpec/K22/K22O04.html}{https://minorplanetcenter.net/mpec/K22/K22O04.html}} 
         
         If 2022~NX$_{1}$ is a natural object, its orbit determination (see Table~\ref{elements}) makes it compatible with that of a 
         near-Earth asteroid (NEA) of the Apollo dynamical class. Its most recent orbit determination is shown in Table~\ref{elements}; it 
         is based on 172 observations with a data-arc span of 142 days and it has been retrieved from Jet Propulsion Laboratory's (JPL) 
         Small-Body Database (SBDB)\footnote{\href{https://ssd.jpl.nasa.gov/tools/sbdb\_lookup.html\#/}
         {https://ssd.jpl.nasa.gov/tools/sbdb\_lookup.html\#/}} provided by the Solar System Dynamics Group (SSDG, 
         \citealt{2011jsrs.conf...87G,2015IAUGA..2256293G}).\footnote{\href{https://ssd.jpl.nasa.gov/}{https://ssd.jpl.nasa.gov/}} The orbit 
         determination is referred to standard epoch JD 2460000.5 TDB, which is also the origin of time in the calculations. 
%
%
      \begin{table}
       \centering
       \fontsize{8}{12pt}\selectfont
       \tabcolsep 0.14truecm
       \caption{\label{elements}Values of the heliocentric Keplerian orbital elements and their respective 1$\sigma$ uncertainties
                of 2022~NX$_{1}$.
               }
       \begin{tabular}{lcc}
        \hline
         Orbital parameter                                 &   & value$\pm$1$\sigma$ uncertainty \\
        \hline
         Semimajor axis, $a$ (AU)                          & = &   1.02192456$\pm$0.00000009     \\
         Eccentricity, $e$                                 & = &   0.02501797$\pm$0.00000006     \\
         Inclination, $i$ (\degr)                          & = &   1.066697$\pm$0.000003         \\
         Longitude of the ascending node, $\Omega$ (\degr) & = & 274.76734$\pm$0.00011           \\
         Argument of perihelion, $\omega$ (\degr)          & = & 169.58306$\pm$0.00011           \\
         Mean anomaly, $M$ (\degr)                         & = &  65.0876$\pm$0.0002             \\
         Perihelion distance, $q$ (AU)                     & = &   0.99635808$\pm$0.00000007     \\
         Aphelion distance, $Q$ (AU)                       & = &   1.04749104$\pm$0.00000009     \\
         Absolute magnitude, $H$ (mag)                     & = &  28.1$\pm$0.8                   \\
        \hline
       \end{tabular}
       \tablefoot{The orbit determination of 2022~NX$_{1}$ is referred to epoch JD 2460000.5 (2023-Feb-25.0) TDB (Barycentric 
                  Dynamical Time, J2000.0 ecliptic and equinox), and it is based on 172 observations with a data-arc span of 142
                  days (solution date, November 21, 2022, 05:23:57 PST). Source: JPL's SBDB.
                 }
      \end{table}
%
%

         The $N$-body simulations carried out to study the orbital evolution of 2022~NX$_{1}$ have been performed using a direct $N$-body 
         code developed by \citet{2003gnbs.book.....A} that is publicly available from the website of the Institute of Astronomy of the 
         University of Cambridge.\footnote{\href{http://www.ast.cam.ac.uk/~sverre/web/pages/nbody.htm}
         {http://www.ast.cam.ac.uk/~sverre/web/pages/nbody.htm}}
         This software applies the Hermite integration scheme formulated by \citet{1991ApJ...369..200M}. Results from this code were 
         discussed in detail by \citet{2012MNRAS.427..728D}. Calculations were carried out in an ecliptic coordinate system with the 
         $X$ axis pointing toward the first point of Aries or vernal equinox and in the ecliptic plane, the $Z$ axis perpendicular to the 
         ecliptic plane and pointing northward, and the $Y$ axis perpendicular to the previous two and defining a right-handed set. Our 
         physical model included the perturbations by the eight major planets, the Moon, the barycenter of the Pluto-Charon system, and the 
         three largest asteroids, (1) Ceres, (2) Pallas, and (4) Vesta. For accurate initial positions and velocities (see, for example, 
         Sect.~\ref{Adata}), we used data from JPL's SSDG {\tt Horizons} online Solar System data and ephemeris computation 
         service,\footnote{\href{https://ssd.jpl.nasa.gov/horizons/}{https://ssd.jpl.nasa.gov/horizons/}} which are based on the DE440/441 
         planetary ephemeris \citep{2021AJ....161..105P}. Most input data were retrieved from JPL's SBDB and {\tt Horizons} using tools 
         provided by the {\tt Python} package {\tt Astroquery} \citep{2019AJ....157...98G} and its {\tt HorizonsClass} 
         class.\footnote{\href{https://astroquery.readthedocs.io/en/latest/jplhorizons/jplhorizons.html}
         {https://astroquery.readthedocs.io/en/latest/jplhorizons/jplhorizons.html}}

         In order to interpret the reflectance spectrum of 2022~NX$_{1}$, we taxonomically classified it using the {\tt Modeling for 
         Asteroids} (M4AST)\footnote{\href{http://spectre.imcce.fr/m4ast/index.php/index/home}
         {http://spectre.imcce.fr/m4ast/index.php/index/home}} online tool \citep{2012A&A...544A.130P}. Then, we compared it to other 
         similar spectra of NEAs.
         
   \section{Results\label{Results}}
      In this section, we use reflectance spectroscopy and $N$-body simulations to determine the nature and actual origin of 2022~NX$_{1}$.

      \subsection{Spectroscopy}
         The visible spectrum of 2022~NX$_{1}$ was obtained on August 6, 2022, 23:40 UTC, using the Optical System for Imaging and Low 
         Resolution Integrated Spectroscopy (OSIRIS) camera spectrograph \citep{2000SPIE.4008..623C,2010ASSP...14...15C} at the 10.4~m Gran 
         Telescopio Canarias (GTC), located at the El Roque de Los Muchachos Observatory (La Palma, Canary Islands). Observations were done 
         under the program GTC23-22A (PI, J.~de~Le{\'o}n). Details on the instrumental setup and the data reductions are provided in 
         Appendix~\ref{Aspectrum}.

%
%
      \begin{figure}
        \centering
         \includegraphics[width=\columnwidth]{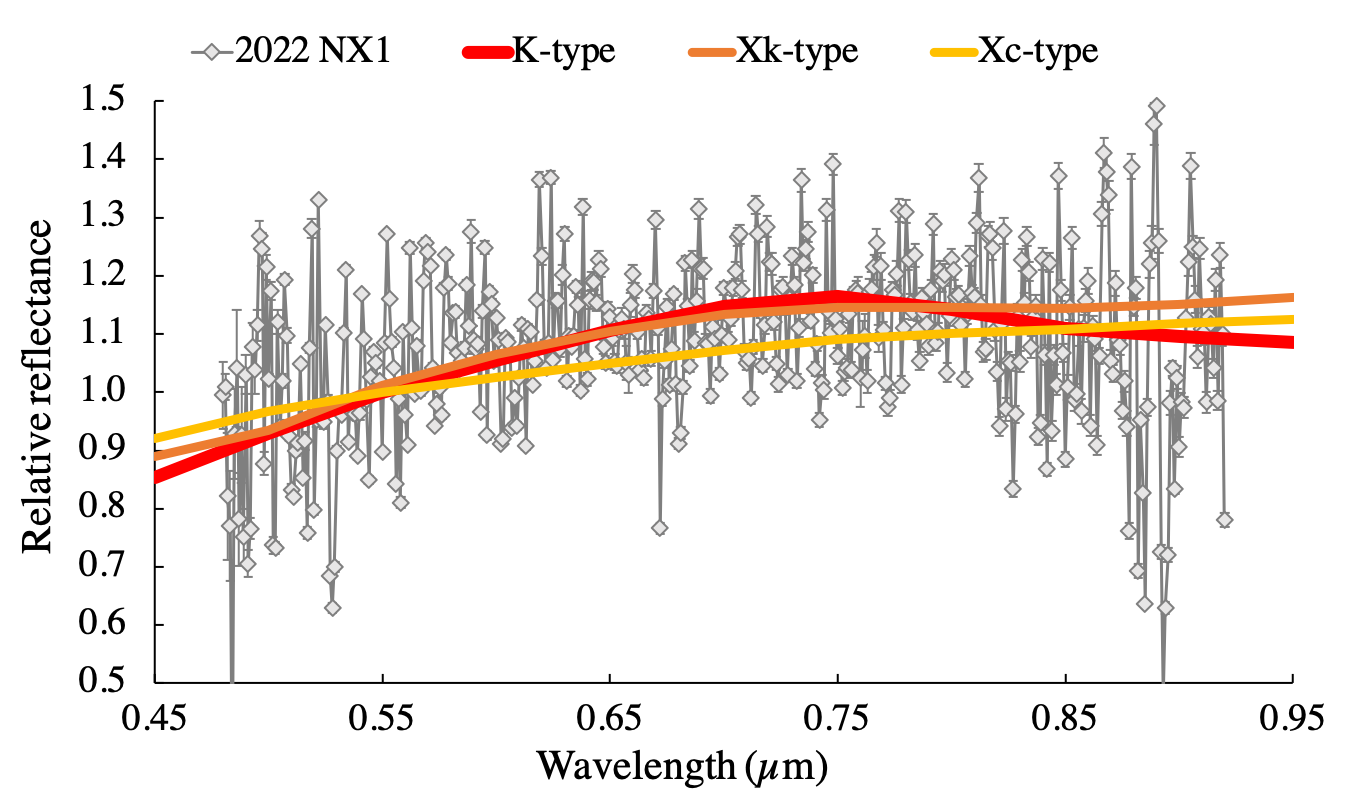}
         \caption{Visible spectrum of 2022~NX$_{1}$ (gray) and its three best taxonomical classifications from the M4AST online tool, in 
                  decreasing order of goodness of fit: K type (in red), Xk type (in orange), and Xc type (in yellow).}
         \label{spectrum}
      \end{figure}
%
%
         The resulting spectrum is shown in Fig.~\ref{spectrum} (gray line). The faintness of the target (apparent visual magnitude $m_V$ = 
         21.2) and the fact that the observations were carried out under less than optimal observing conditions prevented us from using 
         longer exposure times; therefore, extracted individual spectra had a low signal-to-noise ratio (S/N$\sim$30). Nevertheless, it was 
         good enough to allow us to use the M4AST online tool to taxonomically classify it. The tool fits a curve to the data and compares 
         it with the taxons defined by \citet{2009Icar..202..160D} using a $\chi^{2}$ fitting procedure. The three best results are 
         provided, in order of decreasing goodness of fit. In this case, the best fit is with K-type asteroids, followed by Xk-type and 
         Xc-type ones, as shown in Fig.~\ref{spectrum}. Considering the noise in the spectrum, the three classifications can be used to 
         compositionally interpret the spectrum of 2022 NX$_1$, and so, the near-infrared (NIR, up to 2.5~$\mu$m) is needed to actually 
         discern between them: K types have an almost neutral spectral slope in the NIR, with a wide and shallow absorption band at 1~$\mu$m 
         (silicates), while Xk types have a red spectral slope in the NIR and a very slight absorption feature near 0.9 to 1~$\mu$m 
         (intermediate between being carbonaceous like and silicate rich, with lower albedo values), and the Xc types show no feature around 
         1~$\mu$m and present a slightly curved and concave downward spectrum at NIR (carbonaceous like). It is important to remark here 
         that having only the visible spectrum, we can only speculate on the subclasses of the X main taxon, and so, we can only conclude 
         that the object's visible spectrum fits both to a K-type and an X-type taxonomy.
         
%
%
      \begin{figure}
        \centering
         \includegraphics[width=\columnwidth]{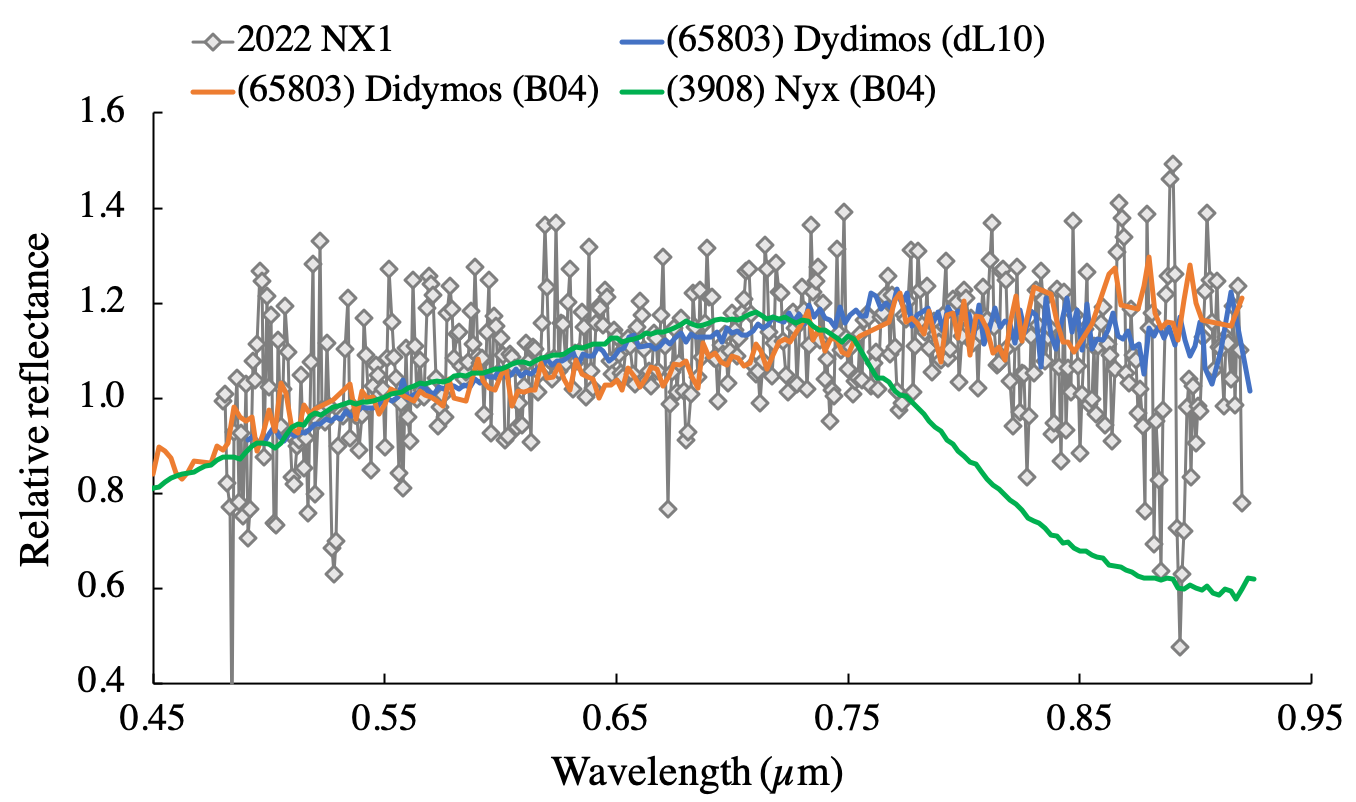}
         \caption{Comparison between the visible spectrum of 2022~NX$_{1}$ (gray) and those of NEA (65803) Didymos from two different 
                  sources: B04, which classifies it as an X-type asteroid (orange, \citealt{2004M&PS...39..351B}), and dL10, which 
                  classifies it as an S-type asteroid from its visible and NIR spectrum (blue, \citealt{2010A&A...517A..23D}). We have also 
                  included the spectrum of one V-type asteroid (green, \citealt{2004M&PS...39..351B}) as a comparison. The spectra have been 
                  normalized to unity at 0.55 $\mu$m. 
                 }
         \label{spectrum3}
      \end{figure}
%
%

         An excellent example of the importance of having the NIR is the case of asteroid (65803) Didymos, which is a target of the Double 
         Asteroid Redirection Test (DART) and Hera missions (see, for example, \citealt{2016P&SS..121...27C}). Following the visible 
         spectrum from \citet{2004M&PS...39..351B}, the object was classified as an Xk-type asteroid (orange line in 
         Figure~\ref{spectrum3}), but later observations that included the NIR (blue line in Figure~\ref{spectrum3}) showed that the object 
         is indeed an S-type asteroid (\citealt{2010A&A...517A..23D}). We have also compared the spectrum of 2022~NX$_{1}$ with the spectra 
         of several artificial objects, including space debris and satellites (see Fig.~\ref{artispectra} in Appendix~\ref{ArtiSpec}). Only 
         one out of six spectra resembles that of 2022~NX$_{1}$, with the remaining five presenting a much redder spectral slope. Therefore, 
         evidence points toward a natural origin for this object. Although we cannot determine the exact taxonomical class, it is evident 
         that the object is not space debris but a NEA of the Apollo dynamical class. Considering its absolute magnitude in 
         Table~\ref{elements} and for typical values of the albedo of K types (range 0.08--0.29 according to \citealt{2011ApJ...741...90M}), 
         2022~NX$_{1}$ may have a size in the range 5 to 15~m. We note that Xk types, the second most likely taxonomy for 2022 NX$_1$, have 
         albedo values in the same range as those of K types, supporting our size estimation.

      \subsection{Orbital evolution}
         Apollo asteroid 2022~NX$_{1}$ experienced a close encounter with our planet at 0.00543~AU on June 26, 2022, well inside the Hill 
         radius of Earth, 0.0098~AU, and at a relative velocity of just 0.96~km~s$^{-1}$. Such a slow close encounter may lead to a 
         temporary capture as in the case of 2020~CD$_{3}$ \citep{2020MNRAS.494.1089D}. On the other hand, the orbit determination in
         Table~\ref{elements} places this object close to the edge of Earth's co-orbital zone that goes from $\sim$0.994~AU to 
         $\sim$1.006~AU (see for example \citealt{2018MNRAS.473.3434D}). Therefore, and in addition to perhaps being temporarily bound to 
         our planet, it may be moving co-orbital to it; in other words, the value of $\lambda_{\rm r}$ may be oscillating instead of 
         circulating in the interval ($-\pi$, $\pi$).         

\begin{figure}
        \centering
         \includegraphics[width=\linewidth]{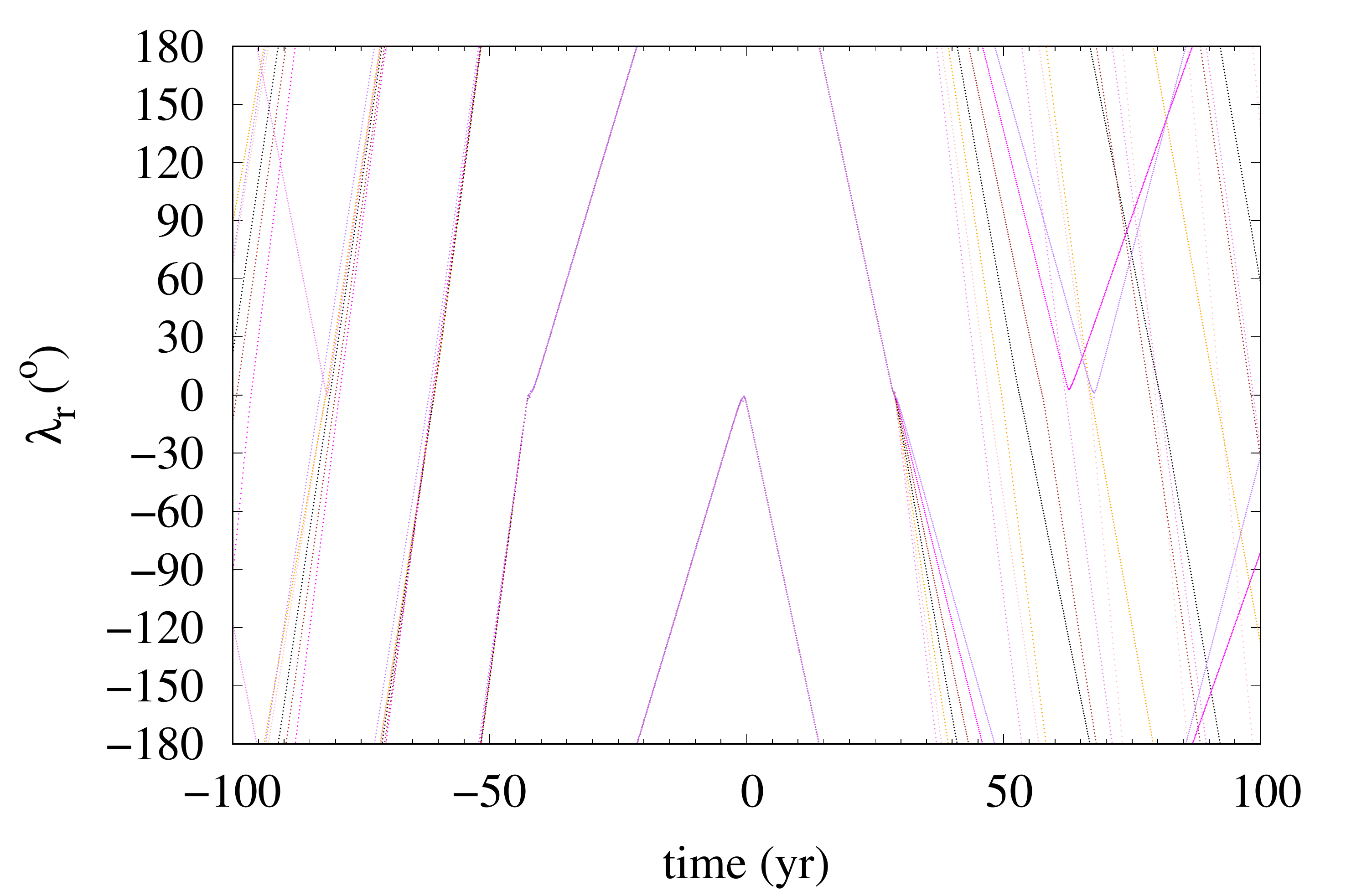}
         \caption{Evolution of the relative mean longitude with respect to Earth, $\lambda_{\rm r}$, of 2022~NX$_{1}$. The time interval 
                  ($-$100, 100)~yr is shown. The figure shows results for the nominal solution (in black) as described by the orbit 
                  determination in Table~\ref{elements} and those of control orbits or clones with Cartesian state vectors (see 
                  Appendix~\ref{Adata}) separated $+$3$\sigma$ (in brown), $-$3$\sigma$ (in orange), $+$6$\sigma$ (in magenta), $-$6$\sigma$ 
                  (in pink), $+$9$\sigma$ (in purple), and $-$9$\sigma$ (in violet) from the nominal values in Table~\ref{vector2022NX1}. 
                  The output time-step size is 0.1~yr.
                 }
         \label{criticalangle}
      \end{figure}

         Figure~\ref{criticalangle} shows that 2022~NX$_{1}$ is currently co-orbital to Earth and follows a horseshoe path with 
         $\lambda_{\rm r}$ librating about 180\degr. This is strictly true for any control orbit with Cartesian state vectors within 
         $\pm$9$\sigma$ from that of the nominal one. The data in the figure also show that the orbital evolution of this object is chaotic
         as its past becomes somewhat unpredictable for times earlier than 1981 (over 40~yr ago) and the same happens in the future, 
         beyond 2051 (or nearly 30~yr from now). This asymmetry is the result of two close encounters with the Earth-Moon system on January 
         16, 1981, at 0.00417~AU and 1.15~km~s$^{-1}$ and on December 4, 2051, at 0.00303~AU and 1.39~km~s$^{-1}$. Slow and deep encounters 
         may result in temporary captures.     
%
%
%
%

         Figure~\ref{criticalangle} shows that, using the orbit determination in Table~\ref{elements}, 2022~NX$_{1}$ has a very short 
         Lyapunov time, $T_{L}$ (the inverse of the maximum Lyapunov exponent). The Lyapunov time is the characteristic timescale for the 
         exponential divergence of initially close orbits. Figure~\ref{criticalangle} shows that $T_{L}$ is about 40~yr for integrations 
         into the past (30~yr for integrations into the future). However, for the Lyapunov time to reach an asymptotic value, a few thousand 
         orbits are needed (see, for example, \citealt{1992AJ....104.1230L}) and 2022~NX$_{1}$ experiences significant orbital changes on a 
         much shorter timescale. The divergence of nearby post-encounter trajectories observed in Fig.~\ref{criticalangle} for the most 
         recent close approach drives future resonant returns that may result in traversing a gravitational keyhole leading to a collision 
         (see, for example, \citealt{2003A&A...408.1179V,2021AJ....162..277R,2022PSJ.....3..123R}). In fact and as of January 2023, 
         2022~NX$_{1}$ has a non-negligible Earth impact risk for approaches starting early in December of 
         2075.\footnote{\href{https://cneos.jpl.nasa.gov/sentry/details.html\#?des=2022 NX1}
         {https://cneos.jpl.nasa.gov/sentry/details.html\#?des=2022 NX1}}

\begin{figure}
        \centering
         \includegraphics[width=\linewidth]{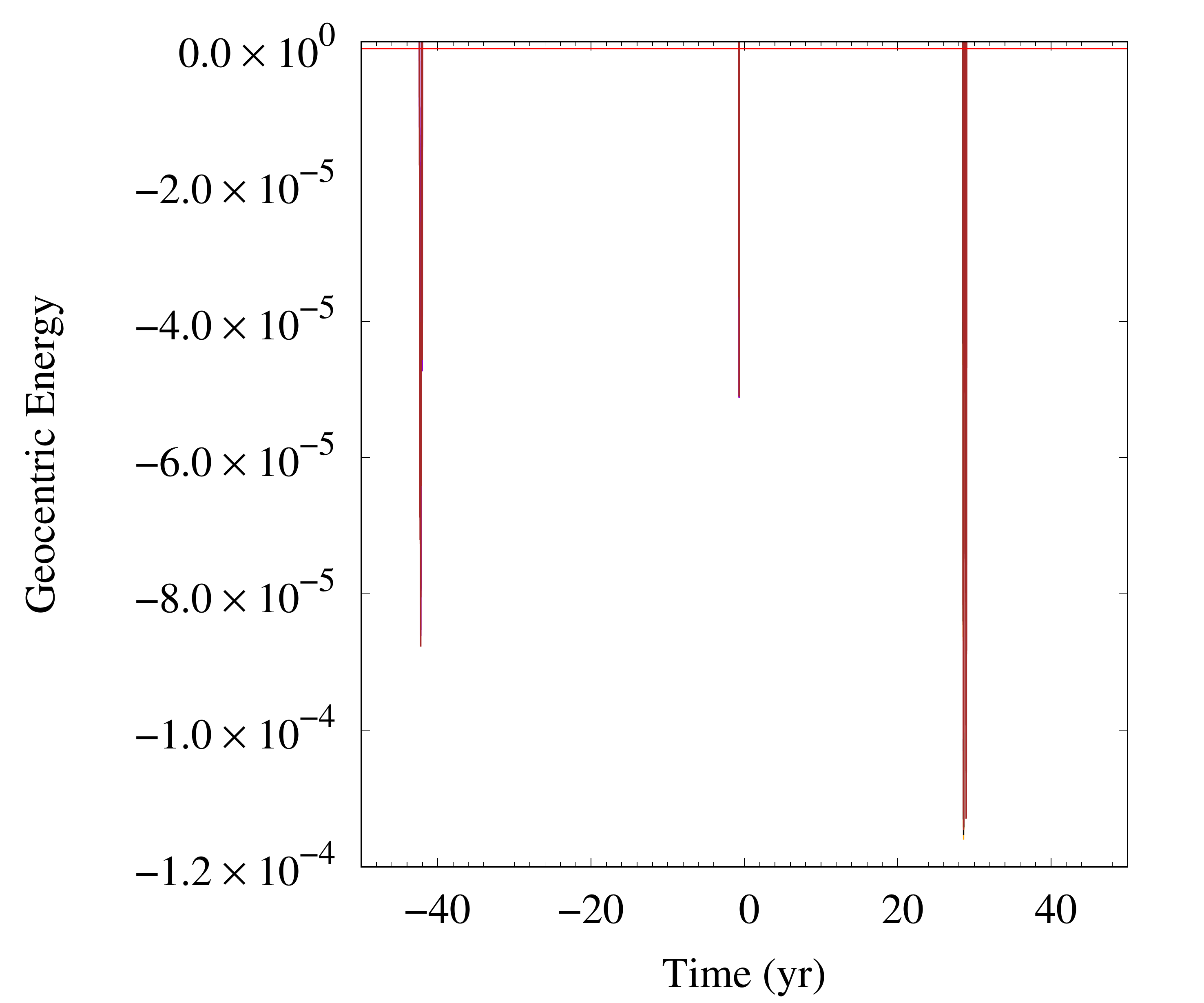}
         \caption{Evolution of the value of the geocentric energy of 2022~NX$_{1}$. Captures happen when the value of the geocentric energy 
                  becomes negative. The unit of energy is such that the unit of mass is 1~$M_{\odot}$, the unit of distance is 1~AU, and the 
                  unit of time is one sidereal year divided by 2$\pi$. The evolution according to the nominal orbit in Table~\ref{elements} 
                  is shown in black, and those of control or clone orbits with Cartesian vectors separated $\pm$3$\sigma$ from the nominal 
                  values in Table~\ref{vector2022NX1} are displayed in orange and brown, respectively.  
                 }
         \label{energy}
      \end{figure}

         Figure~\ref{energy} shows the evolution of the value of the geocentric energy of 2022~NX$_{1}$ in the interval of interest where
         all the control orbits produce consistent results. Capture events are not as deep and long as the ones experienced by 2020~CD$_{3}$ 
         \citep{2020MNRAS.494.1089D}. For this object, we do not observe events leading to moon-moon episodes in which the value of the 
         selenocentric energy of 2022~NX$_{1}$ became negative as documented for 2020~CD$_{3}$ \citep{2020MNRAS.494.1089D}. In this case, 
         capture episodes are similar to those found for 1991~VG (see \citealt{2018MNRAS.473.2939D}). 
%
%
%
%

\begin{figure*}
        \centering
         \includegraphics[width=\linewidth]{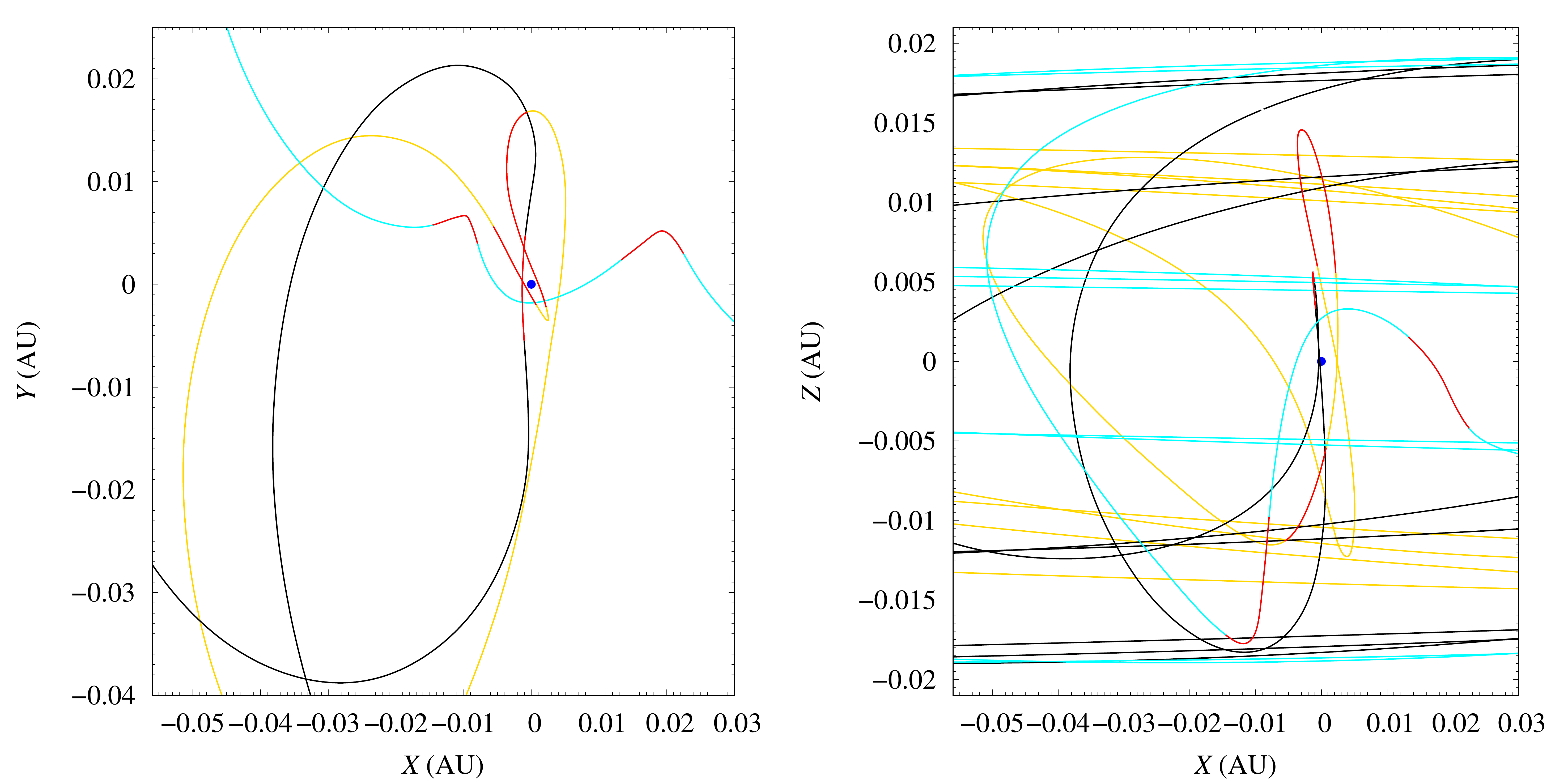}
         \caption{Geocentric trajectories of 2022~NX$_{1}$ during the mini-moon episodes. The flyby in 1981 appears in gold, the 
                  one that occurred in 2022 is in black, and the future flyby in 2051 is in cyan. The part of the trajectory for which the
                  geocentric energy becomes negative is always displayed in red. Earth is represented by a blue dot.
                 }
         \label{geopath}
      \end{figure*}

         Figure~\ref{geopath} shows the geocentric trajectories (in the $XY$ plane, left panel, and the $XZ$ plane, right panel) associated 
         with the capture episodes identified in the time interval of $\pm$70~yr about the reference epoch, namely JD 2460000.5 TDB: January 
         1981, June 2022, and December 2051. These temporary capture episodes are robust and they appear during largely similar time windows 
         for all the control orbits or clones studied here. The figure shows that all the episodes were of the temporarily captured flyby 
         type; in other words, 2022~NX$_{1}$ did not complete even one revolution around our planet while its geocentric energy was 
         negative. Although the most recent mini-moon episode, which lasted from June 11 until July 2 or 21 days, comprised a single 
         temporary capture, the other two events included two captures each. The 1981 episode involved a 98 day-long event (from October 
         1, 1980, until January 7, 1981) and a subsequent, shorter one that lasted 29 days (from January 26, 1981, until February 24, 
         1981); the future 2051 episode will include 63 (from September 4, 2051, until November 6, 2051) and 52 day- (from January 2, 2052, 
         until February 23, 2052) long captures. This is the first time a real object has been found to experience more than one capture 
         event during the same close encounter with Earth. Asteroid 1991~VG may have experienced multiple temporarily captured flyby-type 
         events though always simple \citep{2018MNRAS.473.2939D}, not double as in the case of 2022~NX$_{1}$.
%
%
%
%

\begin{figure}
        \centering
         \includegraphics[width=\linewidth]{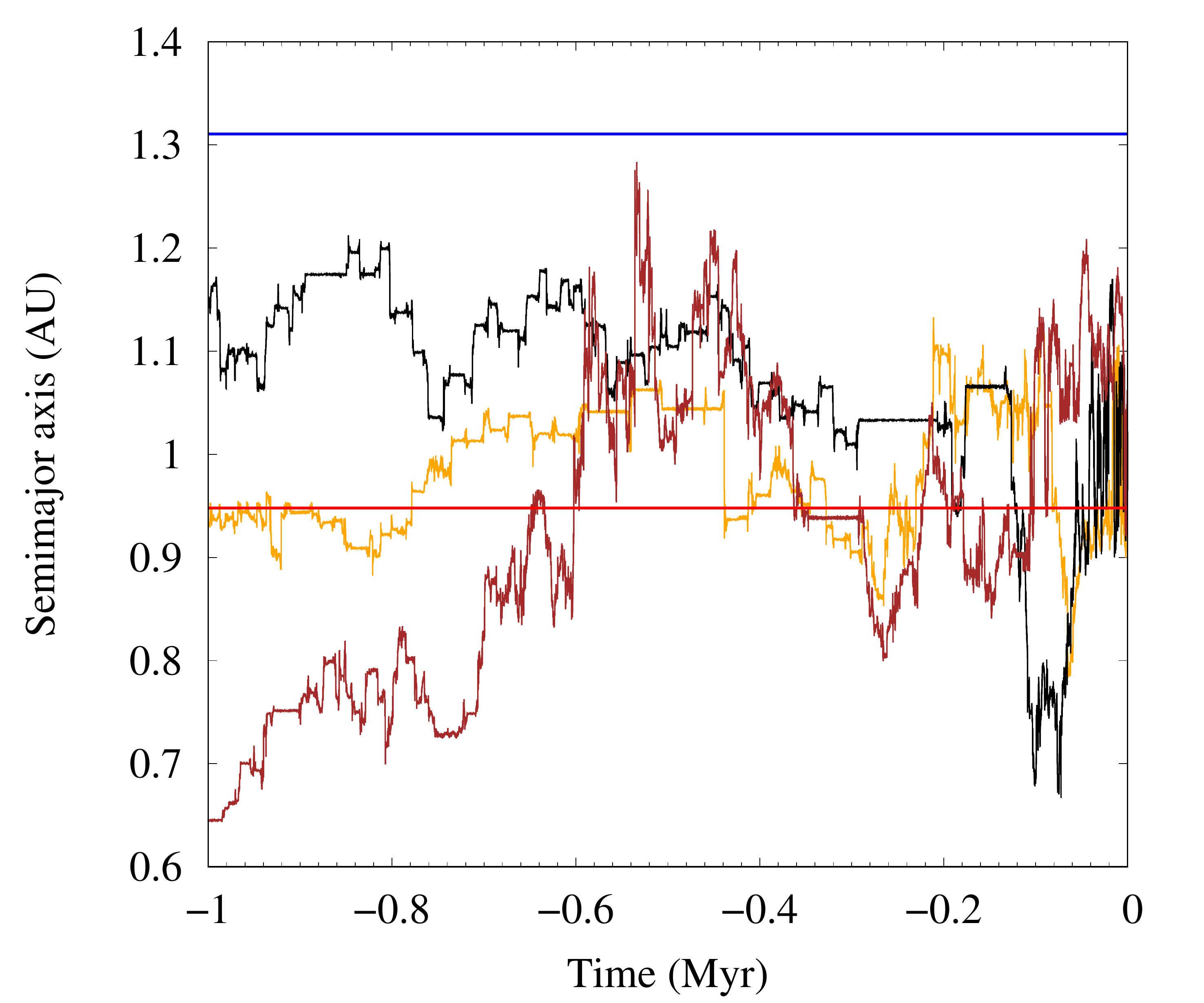}
         \caption{Evolution of the value of the semimajor axis of 2022~NX$_{1}$. The evolution according to the nominal orbit in 
                  Table~\ref{elements} is shown in black and those of control or clone orbits with Cartesian vectors separated 
                  $\pm$3$\sigma$ from the nominal values in Table~\ref{vector2022NX1} are displayed in orange and brown, respectively. In
                  red, we signal the location of the 3:2 external resonance with Venus at 0.9478~AU, and in blue we show the location of the
                  3:2 external resonance with Earth at 1.3104~AU.  
                 }
         \label{semi}
      \end{figure}

         Figure~\ref{criticalangle} shows that the current orbit determination of 2022~NX$_{1}$ does not allow for its orbital evolution to 
         be predicted, beyond a few decades from the current epoch. It is therefore not possible to determine reliably how it may have 
         reached NEA space.  A hint of its possible source can be found in the results plotted in Fig.~\ref{semi}. Although the long-term 
         evolution is chaotic and rather unstable, the value of the semimajor axis remains largely confined within 0.9478~AU (3:2 external 
         resonance with Venus) and 1.3104~AU (3:2 external resonance with Earth). Therefore, and taking into account that 2022~NX$_{1}$ is 
         probably a fragment of a larger NEA, we conjecture that it may have been formed in situ, within the NEA orbital realm during the 
         last few hundred thousand years. On the other hand, we observe multiple (and sometimes lengthy) co-orbital episodes with Earth and 
         relatively brief resonant engagements (when the value of $a$ remains constant).   
%
%
%

   \section{Discussion\label{Discussion}}
      The discovery MPEC of 2022~NX$_{1}$ \citep{2022MPEC....O...04B} suggested that it could have an artificial origin or be lunar ejecta. 
      A number of captured objects ---for example J002E3 and WT1190F--- have been confirmed as space debris thanks to reflectance 
      spectroscopy. However, the spectrum in Fig.~\ref{spectrum} is inconsistent with those of spacecraft materials. On the other hand, 
      Earth quasi-satellite (a co-orbital with $\lambda_{\rm r}$ librating about 0\degr, \citealt{2016MNRAS.462.3441D}) 469219 Kamo`oalewa 
      (2016~HO$_{3}$) could be made of material consistent with what was found in lunar samples \citep{2021ComEE...2..231S} and this 
      provides support to the idea that some of the small bodies inhabiting the near-Earth orbital domain may have their origin on the Moon, 
      likely as lunar ejecta. However, the reflectance spectrum discussed in Sect.~\ref{Results} argues for a natural origin other than the 
      Moon. It is also inconsistent with that of the only other mini-moon with available spectroscopy, 2020~CD$_{3}$, which is a V type
      \citep{2020ApJ...900L..45B} such as asteroid (3908)~Nyx (see green line in Fig.~\ref{spectrum3}). In this context, the K-type spectrum 
      of 2022~NX$_{1}$ argues for a diverse group of natural Earth co-orbitals with different origins and sources that probably reflects the 
      spectral-type distribution of the NEA population (see, for example, \citealt{2019A&A...627A.124P}). K-type asteroids, such as S types, 
      contain rocky silicate minerals and are more common in the inner asteroid belt.

      This is only the fourth time (after 1991~VG, 2006~RH$_{120}$, and 2020~CD$_{3}$) a minor body has been discovered during or right 
      after experiencing an episode of temporary gravitational capture by Earth. As in previous cases, captures are linked to recurrent 
      transient co-orbital motion, in particular of the horseshoe type (see for example \citealt{2018MNRAS.473.2939D,2018MNRAS.473.3434D}). 
      While 2006~RH$_{120}$ and 2020~CD$_{3}$ were identified as temporary captures while still being bound to Earth 
      \citep{2009A&A...495..967K,2020MNRAS.494.1089D}, 1991~VG was not recognized as such until some time later \citep{1997CeMDA..69..119T}. 
      The capture episodes experienced by 1991~VG and 2022~NX$_{1}$ were rather similar (temporarily captured flybys); the same can be said 
      about 2006~RH$_{120}$ and 2020~CD$_{3}$ (temporarily captured orbiters). 

   \section{Summary and conclusions\label{Conclusions}}
      In this Letter, we have presented spectroscopic observations of Earth's horseshoe co-orbital and mini-moon 2022~NX$_{1}$ obtained on 
      August 6, 2022, using the OSIRIS camera spectrograph at the 10.4~m GTC. We used the spectrum to characterize the object. We also 
      carried out direct $N$-body simulations to investigate its orbital evolution. Our conclusions can be summarized as follows.
      \begin{enumerate}
         \item We confirm that 2022~NX$_{1}$ is a natural object but not lunar ejecta.
         \item We find that 2022~NX$_{1}$ has a visible spectrum consistent with that of a K-type asteroid, although it could also be 
               classified as an Xk type. 
         \item We identify two robust, short episodes of the temporarily captured flyby type for 2022~NX$_{1}$ in 1981 and 2022, and 
               predict a third one that will take place in 2051. The temporary capture episodes in 1981 and 2051 include two separate events 
               each.
         \item Considering its absolute magnitude and for typical values of the albedo of K-type asteroids, 2022~NX$_{1}$ may have a size in 
               the range 5 to 15~m that makes it the largest known mini-moon. This result remains valid if it is an Xk-type asteroid. 
         \item We confirm that 2022~NX$_{1}$ inhabits the rim of Earth's co-orbital space, the 1:1 mean-motion resonance, and experiences 
               recurrent co-orbital episodes of the horseshoe type as previous mini-moons did. It is currently following a horseshoe path 
               with respect to Earth.
         \item The current orbit determination of 2022~NX$_{1}$ is not robust enough to reconstruct its past and future orbital evolution 
               beyond $\pm$50~yr from the current epoch.
      \end{enumerate}
      Considering its relatively small size and its probable long-term dynamical evolution into the past, we conjecture that 2022~NX$_{1}$ 
      may have formed via fragmentation within NEA orbital parameter space during the last few hundred thousand years. This preliminary 
      interpretation is based on the currently available data and may change as the orbit determination accuracy improves.

   \begin{acknowledgements}
      We thank an anonymous reviewer for a quick and particularly constructive report. RdlFM and CdlFM thank S.~J. Aarseth for providing one 
      of the codes used in this research and A.~I. G\'omez de Castro for providing access to computing facilities. JdL acknowledges support 
      from the ACIISI, Consejer\'{i}a de Econom\'{i}a, Conocimiento y Empleo del Gobierno de Canarias and the European Regional Development 
      Fund (ERDF) under grant with reference ProID2021010134. JdL also acknowledges financial support from the Spanish Ministry of Science 
      and Innovation (MICINN) through the Spanish State Research Agency, under Severo Ochoa Programme 2020-2023 (CEX2019-000920-S). This 
      work was partially supported by the Spanish `Agencia Estatal de Investigaci\'on (Ministerio de Ciencia e Innovaci\'on)' under grant 
      PID2020-116726RB-I00 /AEI/10.13039/501100011033. Based on observations made with the Gran Telescopio Canarias (GTC), installed at the 
      Spanish Observatorio del Roque de los Muchachos of the Instituto de Astrof\'{\i}sica de Canarias, on the island of La Palma. This work 
      is partly based on data obtained with the instrument OSIRIS, built by a Consortium led by the Instituto de Astrof\'{\i}sica de 
      Canarias in collaboration with the Instituto de Astronom\'{\i}a of the Universidad Nacional Aut\'onoma de Mexico. OSIRIS was funded by 
      GRANTECAN and the National Plan of Astronomy and Astrophysics of the Spanish Government. In preparation of this Letter, we made use of 
      the NASA Astrophysics Data System, the ASTRO-PH e-print server, and the MPC data server. 
   \end{acknowledgements}

   \bibliographystyle{aa}

\begin{thebibliography}{}
      \bibitem[Aarseth(2003)]{2003gnbs.book.....A} Aarseth, S.~J. 2003,
              Gravitational N-Body Simulations
              (Cambridge: Cambridge University Press), 27
      \bibitem[Astropy Collaboration et al.(2013)]{2013A&A...558A..33A} Astropy Collaboration, Robitaille, T.~P., Tollerud, E.~J., et al.\ 2013,
              \aap, 558, A33
      \bibitem[Astropy Collaboration et al.(2018)]{2018AJ....156..123A} Astropy Collaboration, Price-Whelan, A.~M., Sip{\H{o}}cz, B.~M., et al.\ 2018,
              \aj, 156, 123
      \bibitem[Bacci et al.(2022)]{2022MPEC....O...04B} Bacci, P., Maestripieri, M., Grazia, M.~D.\ 2022,
              Minor Planet Electronic Circulars, 2022-O04
      \bibitem[Binzel et al.(2004)]{2004M&PS...39..351B} Binzel, R. P., Perozzi, E., Rivkin, A. S., et al.\ 2004, Meteorit. Planet. Sci., 39, 351
      \bibitem[Bolin et al.(2014)]{2014Icar..241..280B} Bolin, B., Jedicke, R., Granvik, M., et al.\ 2014, 
              \icarus, 241, 280
      \bibitem[Bolin et al.(2020)]{2020ApJ...900L..45B} Bolin, B.~T., Fremling, C., Holt, T.~R., et al.\ 2020, 
              \apjl, 900, L45
      \bibitem[Buzzoni et al.(2019)]{2019AdSpR..63..371B} Buzzoni, A., Altavilla, G., Fan, S., et al.\ 2019, 
              Advances in Space Research, 63, 371
      \bibitem[Cano et al.(2019)]{neosst1} Cano, J.~L., Ceccaroni, M., Faggioli, L., et al.\ 2019,
              ESA's Activities on the Boundaries between NEO and Debris Detection, in 1st NEO and Debris Detection Conference,
              ed.\ T. Flohrer, R. Jehn, \& F. Schmitz (ESA Space Safety Programme Office Publishing), 470
      \bibitem[Carusi \& Valsecchi(1979)]{1979RSAI...22..181C} Carusi, A. \& Valsecchi, G.~B.\ 1979, 
              Riunione della Societa Astronomica Italiana, 22, 181
      \bibitem[Cepa et al.(2000)]{2000SPIE.4008..623C} Cepa, J., Aguiar, M., Escalera, V.~G., et al.\ 2000, 
              \procspie, 4008, 623
      \bibitem[Cepa(2010)]{2010ASSP...14...15C} Cepa, J.\ 2010, 
              Astrophysics and Space Science Proceedings, 14, 15
      \bibitem[Cheng et al.(2016)]{2016P&SS..121...27C} Cheng, A.~F., Michel, P., Jutzi, M., et al.\ 2016, 
              \planss, 121, 27
      \bibitem[Cowardin et al.(2021)]{2021JAnSc..68.1186C} Cowardin, H.~M., Hostetler, J.~M., Murray, J.~I., et al.\ 2021, 
              Journal of the Astronautical Sciences, 68, 1186
      \bibitem[de la Fuente Marcos \& de la Fuente Marcos(2012)]{2012MNRAS.427..728D} de la Fuente Marcos, C. \& de la Fuente Marcos, R.\ 2012, 
              \mnras, 427, 728
      \bibitem[de la Fuente Marcos \& de la Fuente Marcos(2016)]{2016MNRAS.462.3441D} de la Fuente Marcos, C. \& de la Fuente Marcos, R.\ 2016, 
              \mnras, 462, 3441
      \bibitem[de la Fuente Marcos \& de la Fuente Marcos(2018a)]{2018MNRAS.473.2939D} de la Fuente Marcos, C. \& de la Fuente Marcos, R.\ 2018a, 
              \mnras, 473, 2939
      \bibitem[de la Fuente Marcos \& de la Fuente Marcos(2018b)]{2018MNRAS.473.3434D} de la Fuente Marcos, C. \& de la Fuente Marcos, R.\ 2018b, 
              \mnras, 473, 3434
      \bibitem[de la Fuente Marcos \& de la Fuente Marcos(2020)]{2020MNRAS.494.1089D} de la Fuente Marcos, C. \& de la Fuente Marcos, R.\ 2020, 
              \mnras, 494, 1089 
      \bibitem[de la Fuente Marcos \& de la Fuente Marcos(2022)]{2022RNAAS...6..160D} de la Fuente Marcos, C. \& de la Fuente Marcos, R.\ 2022, 
              Research Notes of the American Astronomical Society, 6, 160
      \bibitem[de Le{\'o}n et al.(2010)]{2010A&A...517A..23D} de Le{\'o}n, J., Licandro, J., Serra-Ricart, M., et al.\ 2010, 
              \aap, 517, A23
      \bibitem[DeMeo et al.(2009)]{2009Icar..202..160D} DeMeo, F., Binzel, R. P., Slivan, S. M., et al.\ 2009, 
              \icarus, 202, 160
      \bibitem[Fedorets et al.(2017)]{2017Icar..285...83F} Fedorets, G., Granvik, M., \& Jedicke, R.\ 2017, 
              \icarus, 285, 83
      \bibitem[Fedorets et al.(2020)]{2020AJ....160..277F} Fedorets, G., Micheli, M., Jedicke, R., et al.\ 2020, 
              \aj, 160, 277
      \bibitem[Ginsburg et al.(2019)]{2019AJ....157...98G} Ginsburg, A., Sip{\H{o}}cz, B.~M., Brasseur, C.~E., et al.\ 2019, 
              \aj, 157, 98
      \bibitem[{{Giorgini}(2011)}]{2011jsrs.conf...87G} {Giorgini}, J. 2011,
              in Journ\'ees Syst\`emes de R\'ef\'erence Spatio-temporels 2010,
              ed. N.~{Capitaine}, 87--87
      \bibitem[Giorgini(2015)]{2015IAUGA..2256293G} Giorgini, J.~D.\ 2015,
              IAUGA, 22, 2256293
      \bibitem[Granvik et al.(2012)]{2012Icar..218..262G} Granvik, M., Vaubaillon, J., \& Jedicke, R.\ 2012, 
              \icarus, 218, 262
      \bibitem[Greenstreet et al.(2012)]{2012Icar..217..355G} Greenstreet, S., Ngo, H., \& Gladman, B., 2012,
              \icarus, 217, 355
      \bibitem[Jorgensen(2000)]{2000PhDT........43J} Jorgensen, K.~M.\ 2000, 
              Ph.D. Thesis, University of Colorado, Boulder
      \bibitem[Jorgensen et al.(2003)]{2003DPS....35.3602J} Jorgensen, K., Rivkin, A., Binzel, R., et al.\ 2003, 
              American Astronomical Society, DPS meeting \#35, id.36.02
      \bibitem[Jorgensen et al.(2004)]{2004AdSpR..34.1021J} Jorgensen, K., Africano, J., Hamada, K., et al.\ 2004, 
              Advances in Space Research, 34, 1021
      \bibitem[Kwiatkowski et al.(2009)]{2009A&A...495..967K} Kwiatkowski, T., Kryszczy{\'n}ska, A., Poli{\'n}ska, M., et al.\ 2009, 
              \aap, 495, 967
      \bibitem[Lecar et al.(1992)]{1992AJ....104.1230L} Lecar, M., Franklin, F., \& Murison, M.\ 1992, 
              \aj, 104, 1230
      \bibitem[Licandro et al.(2019)]{2019A&A...625A.133L} Licandro, J., de la Fuente Marcos, C., de la Fuente Marcos, R., et al.\ 2019, 
              \aap, 625, A133
      \bibitem[Mainzer et al.(2011)]{2011ApJ...741...90M} Mainzer, A., Grav, T., Masiero, J., et al.\ 2011, 
              \apj, 741, 90
      \bibitem[Makino(1991)]{1991ApJ...369..200M} Makino, J.\ 1991,
              \apj, 369, 200
      \bibitem[McDowell(2020)]{GCAT} McDowell, J.~C., 2020,
              General Catalog of Artificial Space Objects,
              Release 1.2.12 , {\tt https://planet4589.org/space/gcat}
      \bibitem[Micheli et al.(2018)]{2018Icar..304....4M} Micheli, M., Buzzoni, A., Koschny, D., et al.\ 2018, 
              \icarus, 304, 4
      \bibitem[Morais \& Morbidelli(2002)]{2002Icar..160....1M} Morais, M.~H.~M. \& Morbidelli, A.\ 2002, 
              \icarus, 160, 1
      \bibitem[Murray \& Dermott(1999)]{1999ssd..book.....M} Murray, C.~D., \& Dermott, S.~F.\ 1999,
              Solar System Dynamics
              (Cambridge: Cambridge University Press)
      \bibitem[Park et al.(2021)]{2021AJ....161..105P} Park, R.~S., Folkner, W.~M., Williams, J.~G., et al.\ 2021, 
              \aj, 161, 105 
      \bibitem[Popescu et al.(2012)]{2012A&A...544A.130P} Popescu, M., Birlan, M., \& Nedelcu, D.~A.\ 2012, 
              \aap, 544, A130
      \bibitem[Popescu et al.(2019)]{2019A&A...627A.124P} Popescu, M., Vaduvescu, O., de Le{\'o}n, J., et al.\ 2019, 
              \aap, 627, A124
      \bibitem[Reddy et al.(2022)]{2022PSJ.....3..123R} Reddy, V., Kelley, M.~S., Dotson, J., et al.\ 2022, 
              PSJ, 3, 123
      \bibitem[Roa et al.(2021)]{2021AJ....162..277R} Roa, J., Farnocchia, D., \& Chesley, S.~R.\ 2021, 
              \aj, 162, 277
      \bibitem[Schildknecht(2007)]{2007A&ARv..14...41S} Schildknecht, T.\ 2007, 
              \aapr, 14, 41
      \bibitem[Sharkey et al.(2021)]{2021ComEE...2..231S} Sharkey, B.~N.~L., Reddy, V., Malhotra, R., et al.\ 2021, 
              Communications Earth and Environment, 2, 231
      \bibitem[Tancredi(1997)]{1997CeMDA..69..119T} Tancredi, G.\ 1997, 
              Celestial Mechanics and Dynamical Astronomy, 69, 119
      \bibitem[Valsecchi et al.(2003)]{2003A&A...408.1179V} Valsecchi, G.~B., Milani, A., Gronchi, G.~F., et al.\ 2003, 
              \aap, 408, 1179
      \bibitem[Vananti et al.(2017)]{2017AdSpR..59.2488V} Vananti, A., Schildknecht, T., \& Krag, H.\ 2017, 
              Advances in Space Research, 59, 2488
      \bibitem[Watson(2016)]{Watson2016} Watson, T.\ 2016, 
              \nat, 19162
   \end{thebibliography}

   \begin{appendix}
      \section{Spectroscopic observations and data reduction\label{Aspectrum}}
         We used the OSIRIS camera spectrograph at the 10.4~m GTC. The OSIRIS detector is a mosaic of two Marconi 2048$\times$4096 pixel 
         CCDs, with a plate scale of 0.127"/pixel that provides a field of view of 7.8'$\times$7.8'. The standard operation mode of the 
         instrument uses a 2$\times$2 binning. We used the R300R grism that covers a wavelength range from 0.48 to 0.92~$\mu$m, with a 
         dispersion of 7.74~\AA/pixel for a 0.6" slit. Weather conditions during the observations were less than optimal. Although there 
         were no clouds, the seeing was variable, ranging from 1.0" to 1.5". We therefore used the 1.2" slit, oriented to the parallactic 
         angle, and with the tracking of the telescope at a set rate matching the proper motion of the asteroid. We obtained two consecutive 
         spectra of 300~s of exposure time each, at an airmass of 1.15, offsetting the telescope 10" in the slit direction between the 
         spectra. To obtain the reflectance spectra of the asteroid, we also observed two solar analog stars (Landolt SA 110-361 and SA 
         102-1081), using the same instrumental configuration as for the asteroid, and at a similar airmass. In the case of the stars, we 
         obtained three individual spectra, also offsetting the telescope in the slit direction by 10" between individual spectra. Spectral 
         images of the asteroid and the solar analog stars were bias and flat-field corrected. The 2D spectra were background subtracted and 
         collapsed to 1D by adding all the flux within an aperture (typically defined as the distance from the center of the spatial profile 
         where the intensity is 10\% of the peak intensity). One-dimensional spectra were then wavelength calibrated using Xe+Ne+HgAr arc 
         lamps. We added the two asteroid spectra and averaged, for each solar analog, their corresponding individual spectra. Then, as a 
         final step, we divided the spectrum of the asteroid by the spectrum of each solar analog star, and averaged the two resulting 
         ratios to compute values and error bars. That is the final spectrum shown in Fig.~\ref{spectrum}.

      \section{Reflectance spectra of artificial objects\label{ArtiSpec}}
\begin{figure}
        \centering
         \includegraphics[width=\columnwidth]{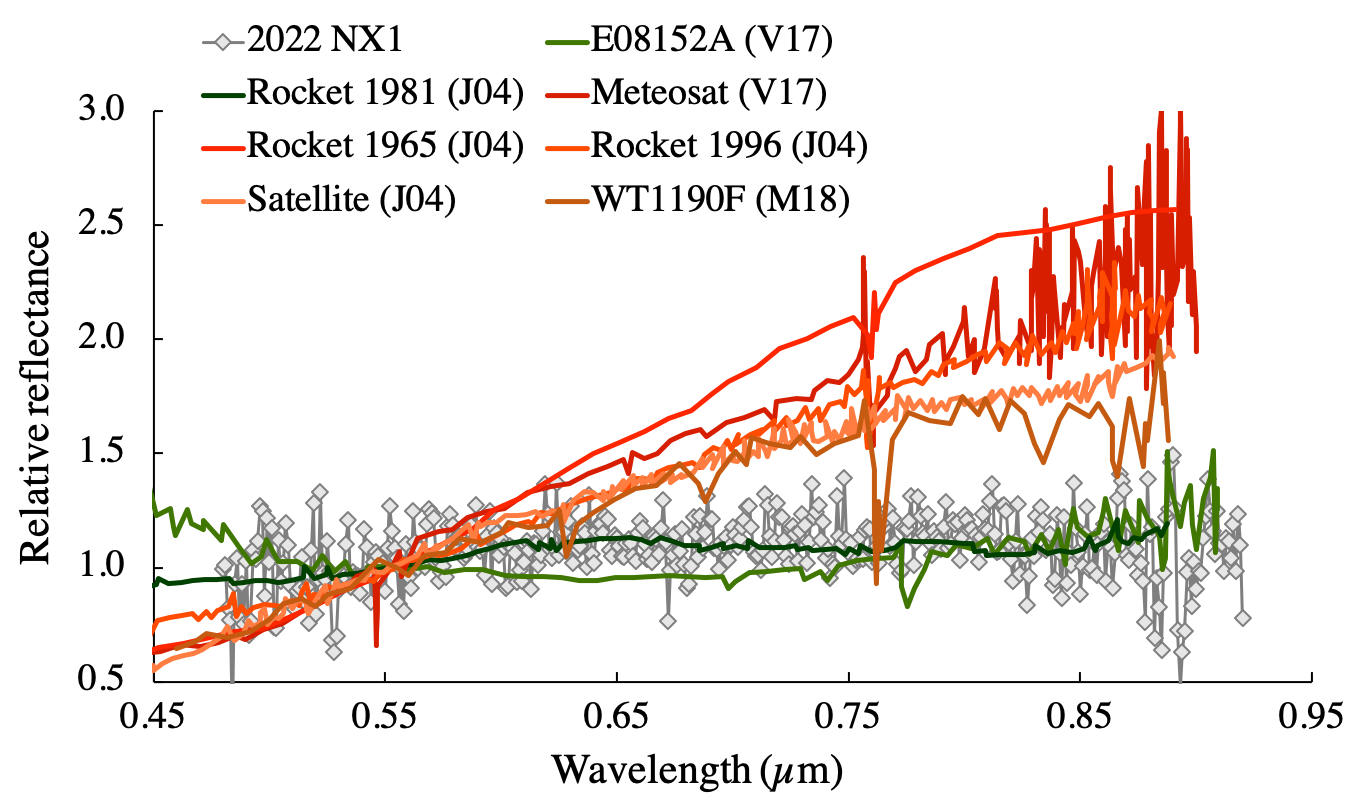}
         \caption{Visible spectrum of 2022~NX$_{1}$ compared with visible spectra of different artificial objects, including rocket bodies, 
                  satellites, and space debris, from several sources: J04 \citep{2004AdSpR..34.1021J}; V17 \citep{2017AdSpR..59.2488V}; and 
                  M18 \citep{2018Icar..304....4M}. The spectra have been normalized to unity at 0.55 $\mu$m.
                 }
         \label{artispectra}
      \end{figure}

         We have collected published spectra of several artificial objects and put them together with our visible spectrum of 2022~NX$_{1}$ 
         in Fig.~\ref{artispectra}. We included several rocket bodies that launched in different years (Rocket 1965, 1981, and 1996), and a 
         satellite, from \citet{2004AdSpR..34.1021J}; the Meteosat satellite and the debris object E08152A in an elliptical geosynchronous 
         equatorial orbit, from \citet{2017AdSpR..59.2488V}; and the artificial object WT1190F, from \citet{2018Icar..304....4M}. As it can 
         be seen, the majority of artificial objects present visible spectra with a much redder spectral slope compared to that of 
         2022~NX$_{1}$. We have found only one rocket body (Rocket 1981) with a similar spectrum to that of 2022~NX$_{1}$.
%
%
%
%

         NEA 2022~NX$_{1}$ experienced a close approach to Earth at 0.004~AU on January 16, 1981. The artificial satellites 
         Kosmos~1238\footnote{\href{https://nssdc.gsfc.nasa.gov/nmc/spacecraft/display.action?id=1981-003A}
         {https://nssdc.gsfc.nasa.gov/nmc/spacecraft/display.action?id=1981-003A}} and
         Kosmos~1239\footnote{\href{https://nssdc.gsfc.nasa.gov/nmc/spacecraft/display.action?id=1981-004A}
         {https://nssdc.gsfc.nasa.gov/nmc/spacecraft/display.action?id=1981-004A}} were launched from Plesetsk on that day. Kosmos~1238
         was placed in a low-Earth orbit by a two-stage Kosmos-3M rocket and Kosmos~1239 by a three-stage Soyuz-U rocket. Both satellites 
         were successfully placed in nearly polar orbits. The timings of the close approach of 2022~NX$_{1}$ and the launch of Kosmos~1239 
         are somewhat consistent. Rocket 1981 from \citet{2004AdSpR..34.1021J} seems to be made of aluminum in white paint; the Soyuz-U 
         rocket that launched Kosmos~1239 was painted mostly whitish in color. Although there is a series of curious coincidences, we 
         consider it highly unlikely that one of the stages of the Soyuz-U rocket that launched Kosmos~1239 may have been able to escape 
         Earth's gravity to reach a heliocentric (but co-orbital to Earth) and low-inclination orbit, and eventually return for another 
         close approach on June 26, 2022. 

      \section{Input data\label{Adata}}
         Here, we include the barycentric Cartesian state vector of NEA 2022~NX$_{1}$. This vector and its uncertainties have been used to 
         perform the calculations discussed above and to generate the figure that displays the time evolution of the critical angle, 
         $\lambda_{\rm r}$. For example, a new value of the $X$ component of the state vector is computed as $X_{\rm c} = X + \sigma_X \ r$, 
         where $r$ is an univariate Gaussian random number, and $X$ and $\sigma_X$ are the mean value and its 1$\sigma$ uncertainty in  
         Table~\ref{vector2022NX1}.
%
%
     \begin{table}
      \centering
      \fontsize{8}{12pt}\selectfont
      \tabcolsep 0.15truecm
      \caption{\label{vector2022NX1}Barycentric Cartesian state vector of 2022~NX$_{1}$: components and associated 1$\sigma$ uncertainties.
              }
      \begin{tabular}{ccc}
       \hline
        Component                         &   &    value$\pm$1$\sigma$ uncertainty                                 \\
       \hline
        $X$ (AU)                          & = & $-$9.027005168802194$\times10^{-1}$$\pm$1.15841197$\times10^{-7}$  \\
        $Y$ (AU)                          & = &    4.734562771552952$\times10^{-1}$$\pm$2.20475306$\times10^{-7}$  \\
        $Z$ (AU)                          & = & $-$1.563722540015655$\times10^{-2}$$\pm$2.51735401$\times10^{-8}$  \\
        $V_X$ (AU/d)                      & = & $-$8.392947703131247$\times10^{-3}$$\pm$2.25199602$\times10^{-9}$  \\
        $V_Y$ (AU/d)                      & = & $-$1.500620447456039$\times10^{-2}$$\pm$3.23298503$\times10^{-9}$  \\
        $V_Z$ (AU/d)                      & = & $-$1.789369608185863$\times10^{-4}$$\pm$9.62976703$\times10^{-10}$ \\
       \hline
      \end{tabular}
      \tablefoot{Data are referred to epoch JD 2460000.5, which corresponds to 0:00 on February 25, 2023, TDB (J2000.0 ecliptic and equinox). 
                 Source: JPL's {\tt Horizons}.
                }
     \end{table}
%
%

   \end{appendix}

\end{document}